# Single-photon spectroscopy of a single molecule

Y. L. A. Rezus[1,2], S. G. Walt[1], R. Lettow[1,3], A. Renn[1], G. Zumofen[1], S. Götzinger[1], and

V. Sandoghdar[1,4]

1 Laboratory of Physical Chemistry, ETH Zurich, 8093 Zurich, Switzerland

2 Present address: FOM Institute for Atomic and Molecular Physics (AMOLF), 1098 XG Amsterdam, The Netherlands

3 Present address: Industry Division Leica Microsystems AG, 9435 Heerbrugg, Switzerland

4 Present address: Max Planck Institute for the Science of Light & Department of Physics, University of Erlangen-Nuremberg, 91058 Erlangen, Germany

Abstract:

Exploring the interaction of light and matter at the ultimate limit of single photons and single emitters is of great interest both from a fundamental point of view and for emerging applications in quantum engineering. However, the difficulty of generating single photons with specific wavelengths, bandwidths and brightness as well as the weak interaction probability of a single photon with an optical emitter pose a formidable challenge toward this goal. Here, we demonstrate a general approach based on the creation of single photons from a single emitter and their use for performing spectroscopy on a second emitter situated at a distance. Although we used organic molecules as emitters, our strategy is readily extendable to other material systems such as quantum dots and color centers. Our work ushers in a new line of experiments that provide access to the coherent and nonlinear couplings of few emitters and few propagating photons.

Interaction of a propagating photon with an atom or molecule has been a corner stone of quantum physics[1] and plays an elementary role in any optical phenomenon, but its discussions have been limited to "gedanken" experiments. By the end of the last century, a number of techniques were devised for experimental studies of single material particles such as ions, atoms, molecules, and quantum dots, and generation of single-photon streams became of age. In the wake of these developments, several theoretical works have shown that single photons can be directly and perfectly absorbed or reflected by single emitters without the need for microresonators[2-6]. These proposals set the ground for optical signal processing at the nanometer scale, where a quantum emitter can be used as the smallest elementary medium for achieving optical nonlinearity, and act as transistor, phase shifter, or memory at the single-photon level[7-10]. Some of the concepts have been recently examined in the laboratory using laser beams[11-14], however single-photon investigations have been elusive because of the stringent requirements on the performance of the single-photon source and the need for a very efficient emitter-photon interfacing. In this paper we report on a versatile strategy that uses single photons emitted from one molecule to perform spectroscopy on a second molecule placed at a distance of several meters.

A single quantum emitter can generate up to one photon per spontaneous emission cycle, corresponding to a radiated power of 0.03-3 nW if one assumes a typical fluorescence lifetime of the order of 0.1-10 ns for dipole-allowed transitions. Although recent reports have shown that a very large fraction of this fluorescence can be collected[15], losses in the detection path commonly limit the usable power of single-photon sources to less than pW. Detection of this weak light on top of residual background fluorescence and detector noise was, in fact, the central challenge of single-molecule spectroscopy, which was surmounted in the 1990s[16]. Considering that in these experiments one usually relies on lasers with mW or higher power, detection of one emitter with single photons generated by another might appear as a daunting experimental task. To achieve this, one has to both collect the photons from the "source" particle and funnel them to the "target" particle very efficiently. Furthermore, the frequency of the single-photon source should be tunable with respect to that of the target transition if one is to perform spectroscopy.

Three conceptual points have made it possible for us to reach these goals. First, the extinction cross section of a single two-level emitter is given by $\sigma = \frac{3\lambda^2}{2\pi}\frac{\gamma_0}{\gamma_{hom}}$ where λ, $\gamma_0$, and $\gamma_{hom}$ are the wavelength, natural linewidth, and homogeneous width of the transition at hand, respectively[17]. Second, if $\gamma_{hom} \approx \gamma_0$ can be met for an emitter, its cross section reduces to about $\lambda^2/2$, which is comparable to the area of a beam focused with high numerical aperture (NA). Finally, the ultimate limit on the signal-to-noise ratio in extinction spectroscopy is set by the intensity shot noise of the excitation beam[18]. Therefore, deciphering an extinction signal of the order of one percent within one second would require a flux larger than about $10^4$ photons per second in a beam that is tightly focusable. This flux can be obtained from a single emitter if high-NA collection optics is used.

In our current experiment we used the organic dye molecule dibenzanthanthrene (DBATT) embedded in organic matrix tetradecane at T=1.4 K. Under these conditions, the photons emitted on the narrow zero-phonon lines (00-ZPL) between the lowest vibrational levels (v=0) of the electronic ground and excited states (see Fig. 1A) have a coherence length that reaches several meters and is only limited by the spontaneous emission rate of the excited state (i.e. $\gamma_0 \approx \gamma_{hom}$). As a result, the 00-ZPL cross section falls short of the optimal value of $3\lambda^2/2\pi$ only by 2-3 times caused by fluorescence to the $v \neq 0$ levels and the phonon wings[12,16].

Figure 1B shows the schematics of our experimental arrangement, where two different samples were mounted in separate cryogenic confocal microscopes. In each sample, the inhomogeneous distribution of the sharp 00-ZPL frequencies ensures that individual DBATT molecules can be excited selectively by tuning the frequency of a narrow-band light source around a wavelength of λ=589 nm.

Commonly, the red-shifted fluorescence of the v=0 level in the excited state to the $v \neq 0$ levels of the ground state is recorded[15]. However, recent developments have shown that single molecules can also be detected via coherent extinction spectroscopy[12]. Figure 2a displays an example of a 9.3% attenuation dip obtained when the frequency of a laser beam was scanned across the 00-ZPL resonance of a molecule with a full width at half-maximum (FWHM) of 20 MHz. In what follows, we will use this molecule as the target molecule.

To produce photons that matched the 00-ZPL frequency of the target molecule ($\omega_t$), we next recorded excitation spectra from the source sample within a few GHz of $\omega_t$. Once a suitable source molecule was found and characterized, it was excited via the 01 transition between the v=0 and v=1 levels of its ground and excited electronic states, respectively (see Fig. 1A). A rapid radiationless decay of the v=1 level in the excited state populates its v=0 level, which then decays radiatively. By using suitable spectral filters, we isolated the narrow-band spontaneous emission on the 00-ZPL from the Stokes-shifted fluorescence to $v \neq 0$ levels. A Hanbury-Brown and Twiss correlation measurement on this emission is presented in Fig. 2B and shows a clear reduction of the normalized second-order autocorrelation function $g^{(2)}(\tau = 0)$ below 0.5, verifying its single-photon character[17]. We emphasize that although the generated photons do not follow a triggered train, the resulting stream consists of individual photon pulses separated in time.

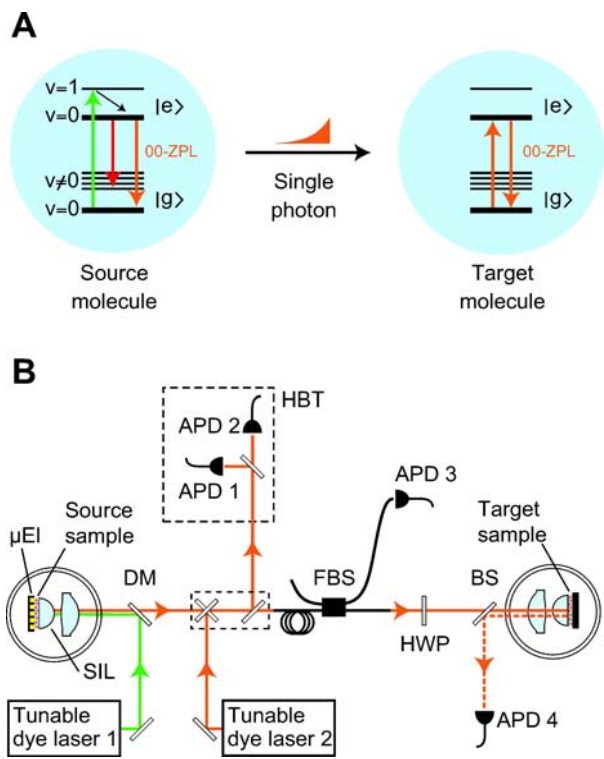

**Figure 1 Experimental configuration. (A)** Schematic view of the free-space connection between two molecules via a single photon. The energy-level diagrams of the source and target molecules as well as various relevant excitation and emission channels are shown. **(B)** Experimental setup: Two separate cryogenic microscopes house the source and target samples. Two ring dye lasers are used to probe the 01 and 00 transitions of the molecules. µEl: microelectrodes; SIL: solid immersion lens; DM: dichroic mirror; HBT: Hanbury-Brown Twiss correlator; FBS: polarization-maintaining single-mode fiber-coupled beam splitter; HWP: half-wave plate; BS: 50:50 beam splitter; APD: avalanche photodiode. A bandpass filter at 589 is not shown.

The schematics of our experimental setup is shown in Fig. 1B. In both the source and target microscopes, we used a combination of aspherical and solid-immersion lenses for achieving a high NA in collection and excitation[12]. Two continuous-wave tunable dye lasers were employed for addressing the 00 and 01 transitions of DBATT, which are typically separated in wavelength by about 9 nm. Avalanche photodiodes (APD) were used as detectors in various stations. A polarization-maintaining single-mode optical fiber was used to facilitate the alignment of the optical axes of the two microscopes and to ensure a good spatial mode quality. An integrated beam splitter with variable transmission allowed us to probe the light in the fiber. In this work, we examined the extinction of the incident beam on APD4 after it was reflected from a gold mirror deposited on the back of the target sample[19]. This arrangement can be regarded as a folded transmission experiment.

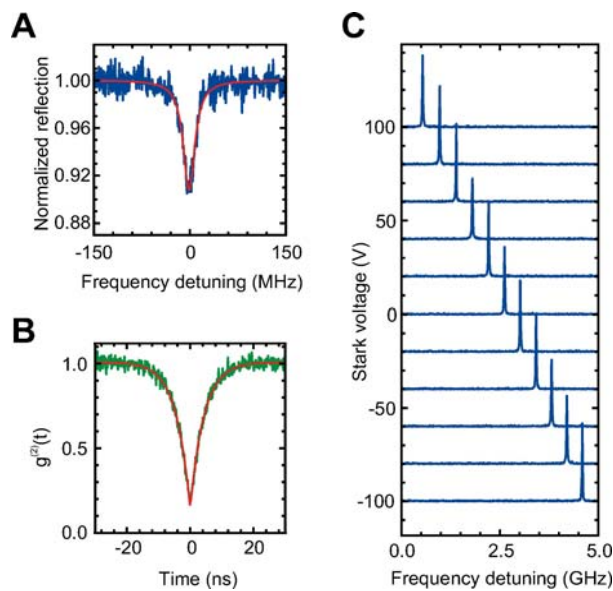

**Figure 2 Characterization of the target and source molecules. (A)** The intensity of a laser beam reflected by the target sample as a function of laser frequency detuning. The red line represents a Lorentzian fit with FWHM of 20 MHz and amplitude of 9.3%. **(B)** Intensity autocorrelation of the light emitted by the source molecule. The red line represents a fit to a double-sided exponential with a time constant of 4.2 ns. **(C)** Zero-phonon fluorescence excitation spectra of the source molecule for various voltages applied to the microelectrodes.

In order to perform spectroscopy on the target molecule with the photon stream generated by the source molecule, the 00-ZPL transition frequency of one has to be tuned through the other. We achieved this by applying a voltage to the interdigitated microelectrodes deposited on the source sample substrate and shifting the molecular resonance via Stark effect. Figure 2C displays a series of 00-ZPL spectra obtained at different voltages by recording the Stokes-shifted fluorescence of the source molecule on APD1. Finally, we aligned the polarization of the single-photon stream with the transition dipole moment of the target molecule using a half-wave plate; we did not correct for a residual ellipticity of up to 10% acquired in the optical path.

The symbols in Fig. 3 present the intensity of the single-photon stream on APD4 as its frequency ($\omega_s$) was scanned through $\omega_t$. We find an attenuation of about 3% and a FWHM of about 60 MHz, in contrast to a dip of 9.3% and FWHM of about 20 MHz obtained in the extinction spectrum of the same molecule recorded under laser excitation (see Fig. 2A). The difference between the two spectra stems from the fact that the linewidth of our dye laser is about 1 MHz, while the single photons generated via a spontaneous emission have a minimum linewidth of 20 MHz. In our case, the experimentally measured rise time of 4.2 ns in $g^{(2)}$ (see Fig. 2B) reveals that the single-photon source has been broadened to 38 MHz. This can be explained by the high depopulation rate of the ground state in the 01 pump process used here. We also note a slight shift of the center frequency, which we attribute to a delay in the data acquisition process when the Stark voltage was incremented. At this point, it is worth emphasizing that the fluorescence signal of the molecule corresponding to the measurement in Fig. 3 amounts to only a few counts per second; the detection of this weak signal would require the very best available APDs and a long integration time[18].

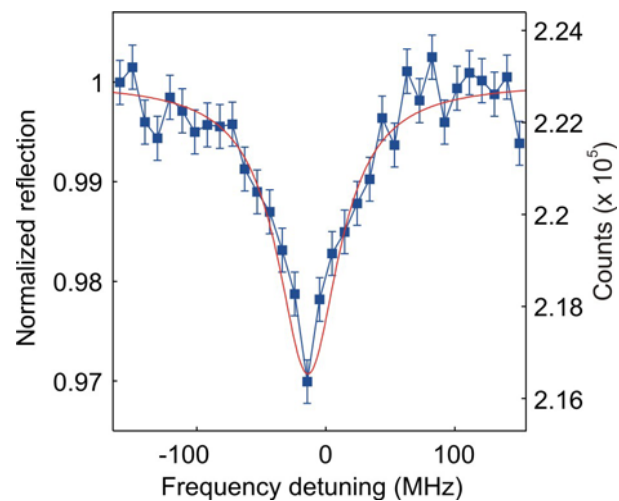

**Figure 3 Extinction spectrum of a single molecule probed by single photons.** Reflection of the single photons emitted by the source molecule from the target molecule as a function of frequency detuning between their 00-ZPLs. The target molecule was the same as the one probed by a laser beam in Fig. 2A. The right-hand axis shows the total number of detected counts after subtracting $2\times10^4$ dark counts per pixel. The left-hand axis displays the normalized intensity of the single-photon stream. The solid curve represents the theoretical calculation (see text for details).

Extinction of a light beam by a sample can be described by considering the addition of the incident and scattered fields[12,20]. To this end, the spectrum in Fig. 3 provides a direct signature of the interference between the probability amplitudes of a single photon traversing the target molecule and scattering from it. To elucidate this process, we have theoretically considered the incidence of a photon wavepacket at central frequency $\omega_s$ and spectral width $\gamma_s$ onto a two-level system with transition frequency $\omega_t=\omega_s$ and a natural linewidth $\gamma_t=\gamma_s$[21,22]. As indicated in Fig. 4A, we took an exponentially rising temporal pulse to emulate the propagation of a photon that is spontaneously emitted by the source molecule with a 1/*e* time of $\tau_s=1/2\pi\gamma_s$ [5,23]. The curves in Figs. 4B and 4C display snap shots of the wavepacket evolution at times $t=2\tau_s$ and $8\tau_s$ after the first encounter of the photon and the target molecule. The reflected components of the photon wavepacket (red curves) display oscillations that signify its interference with the incoming part. The frequency spectrum of the reflected photon is plotted in Fig. 4C and can be associated with the resonance fluorescence (emission) spectrum of the target molecule in the backward half-space. This spectrum is narrower than that of the incident pulse (black) because it results from a temporal convolution of the latter with the exponential response of the molecule during the excitation process, yielding a scattered pulse that is stretched in time.

The theoretical model we have used[21,22] assumes a perfect coupling between the incident light and the emitter, leading to the perfect reflection of a monochromatic and resonant photon[4] (see also supplementary online material). However, if the incident photon has a finite spectral width, as considered above in the condition $\gamma_s=\gamma_t$ , some light is leaked to the transmitted channel. The blue curve in Figs. 4B and 4C show the evolution of the transmitted photon at two different times, and the green curve in Fig. 4C displays a ten-fold zoom of the latter. We find that the transmitted component is fractionated in an approximately exponential pulse that is narrower than the incident one, followed by a flatter envelope. The frequency spectrum corresponding to this temporal behavior is a double-peaked profile that goes to zero on resonance as depicted by the blue curve in Fig. 4D. This spectrum can also be obtained from the subtraction of the incident and reflected spectra. The direct experimental observation of the spectra of the reflected and transmitted channels requires a very strong interaction between the incoming photon and the emitter to ensure a substantial effect of the molecular scattering on top of the incident background field. Future improvements, as discussed below, should make such measurements possible. Furthermore, triggered generation of single photons via pulsed excitation can provide access to the temporal profiles of the reflected and transmitted photons as discussed in Fig. 4C.

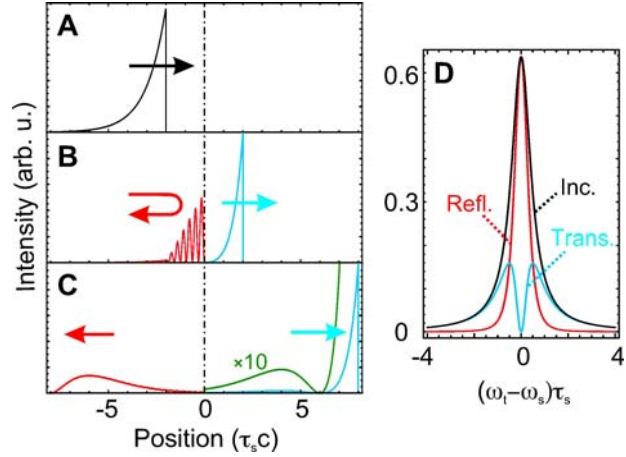

Fig. 4 **Scattering of a photon by a single atom**. (a-c) The electric field intensity distribution of a photon as a function of the position *z* in units of $\tau_s c$. (a) The incident pulse before encountering the emitter. (b) The red and blue curves show the wavepacket intensities in the backward and forward half-spaces at $2\tau_s$ after the first encounter. To visualize the standing wave pattern, we have assumed a small transition frequency $\omega_s = 10/\tau_s$. (c) Same as in (b) but at time $8\tau_s$. The green curve displays a ten-fold zoom of the blue curve. (d) The black, red, and blue profiles plot the frequency spectra associated with the incident, reflected and transmitted photon pulses at long times t>>$\tau_s$, respectively.

In order to reconstruct the extinction spectra obtained in our measurements, we have taken into account the experimentally measured parameters. First, we used $\tau_s = \tau_t = 8$ ns as deduced from the measured FWHM of 20 MHz for the 00-ZPL transition (see Fig. 2A). To consider the spectral broadening of the single photons to a width of 38 MHz mentioned earlier, we introduced an uncertainty of 18 MHz in $\omega_s$. Next, we repeated the calculations for different frequency detunings ($\omega_t - \omega_s$) and plotted the areas under the emission spectra for each case. We then accounted for the experimental extinction efficiency of our system by equating the area under the spectra of Fig. 2A and Fig. 3. These considerations yielded a Lorentzian extinction spectrum with a FWHM of 58 MHz and amplitude of 2.9% (see also supplementary online material), which as displayed by the red curve in Fig. 3, provides an excellent agreement with the experimental spectrum.

In the past few years, clever efforts were embarked for exploring the interaction between single photons and single emitters. For example, single photons generated by a trapped ion were redirected back toward the ion upon reflection from a mirror[24]. Also, correlated photons generated in a down-conversion process have been used to record heralded absorption by a single ion[25]. The experiment described here represents the first demonstration of the direct coupling between two distant quantum emitters via a propagating stream of single photons. While the majority of studies on single quantum emitters rely on the detection of incoherent fluorescence, we have investigated the *coherent* scattering of a single photon by a single molecule. This demonstration sets the stage for achieving strong coherent coupling of two or more emitters separated by much smaller than the coherence length of the photon so that its back and forth scattering from the emitters can couple them in a fashion similar to dipole-dipole coupling[26,27]. To enter this regime, we envision replacing free-space lens coupling by near-field coupling via a nanoguide[9,28,29], which can serve as a bus for connecting many molecules or other quantum emitters within distances of tens or hundreds of micrometers on a chip.

The experimental scheme demonstrated here can be made more efficient by improving the spatial and spectral aspects of photon-emitter interfacing. Here, a modest inhibition of the spontaneous emission of the source molecule, e.g. via plasmonic nanoantennas[30], would render the spectral bandwidth of the single photons narrower than the target transition, improving the coupling by up to a factor of two. Furthermore, one can engineer the radiation pattern of the source by an optical antenna and reach near-unity efficiency in collecting its emitted photons[15]. By reciprocity, such an antenna would also allow nearly perfect mode matching to the radiation pattern of the target molecule and therefore ideal coupling[4]. Implementation of these measures would ensure that each photon interacts with the target emitter with very high probability. This, in turn, paves the way for accessing optical nonlinear effects at the few-photon level[7,8,10] by coupling an emitter to two or more triggered photons that arrive at the same time. In closing, we emphasize that our methodology is not limited to organic molecules and is applicable to other solid-state emitters. Furthermore, it opens the door to a simple far-field optical connection among hybrid systems consisting of quantum dots, color centers, atoms or ions.

We acknowledge the financial support from the Swiss National Science Foundation, ETH Zurich, and the Max Planck Society. Y.L.A.R. acknowledges support from the ETH fellowship program. We also thank I. Gerhardt, G. Wrigge, J. Hwang, M. Pototschnig for their contributions to the very early stages of this project.

**References**

1. A. Einstein, *Annalen der Physik* **322**, 132-148 (1905).
2. J. T. Shen and S. Fan, *Opt. Lett.* **30**, 2001 (2005).
3. D. Pinotsi & A. Imamoglu, *Phys. Rev. Lett.* **100**, 093603 (2008).
4. G. Zumofen, N. M. Mojarad, V. Sandoghdar, M. Agio, *Phys. Rev. Lett.* **101**, 180404 (2008).
5. M. Stobinska, G. Alber, G. Leuchs, *Europhys. Lett.* **86**, 14007 (2009).
6. Y. Wang, J. Minar, L. Sheridan, V. Scarani, Phys. Rev. A **83**, 063842 (2011).
7. P. Domokos, P. Horak, H. Ritsch, *Phys. Rev. A* **65**, 033832 (2002).
8. J.-T. Shen and S. Fan, *Phys. Rev. Lett.* **98**, 153003 (2007).
9. D. E. Chang, A. S. Sorensen, E. A. Demler, M. D. Lukin, *Nature Phys.* **3**, 807 (2007).
10. P. Longo, P. Schmitteckert, K. Busch, *Phys. Rev. Lett.* **104**, 023602 (2010).
11. A. N. Vamivakas, *et al., Nano Lett.* **7**, 2892 (2007).
12. G. Wrigge, I. Gerhardt, J. Hwang, G. Zumofen, V. Sandoghdar, *Nature Phys.* **4**, 60 (2008).
13. J. Hwang, *et al.*, *Nature* **460**, 76 (2009).
14. S. A. Aljunid, *et al., Phys. Rev. Lett.* **103**, 153601 (2009).
15. K. G. Lee, *et al.*, *Nature Photon.* **5**, 166 (2011).
16. T. Basche, W. E. Moerner, M. Orrit, U. P. Wild, *Single-Molecule Optical Detection, Imaging and Spectroscopy* (VCH Weinheim, 1996).
17. R. Loudon, *The Quantum Theory of Light*, Third Edition (Oxford University Press, 2000).
18. G. Wrigge, J. Hwang, I. Gerhardt, G. Zumofen, V. Sandoghdar, *Opt. Express* **16**, 17358 (2008).
19. We neglect standing-wave effects here because the mirror is far enough from the tight focus of the incident beam so that the overlap between the reflected and incident lights does not cause significant modulations. Furthermore, such an effect is automatically contained in the overall interaction efficiency, which is common to the spectra recorded by single-photon and laser excitations.
20. J. D. Jackson, J. D. *Classical Electrodynamics*. (John Wiley & Sons, 1998).
21. U. Dorner, P. Zoller, *Phys. Rev. A* **66**, 023816 (2002).
22. V. Buzek, G. Drobny, E. G. Kim, M. Havukainen, P. L. Knight, *Phys. Rev. A* **60**, 582 (1999).
23. M. O. Scully, S. Zubairy, *Quantum Optics* (Cambridge University Press 1997).
24. J. Eschner, C. Raab, F. Schmidt-Kaler, R. Blatt, *Nature* **413**, 495 (2001).
25. N. Piro, *et al, Nature Phys.* **7**, 17 (2011).
26. C. Hettich, *et al.*, *Science* **298**, 385 (2002).
27. S. Rist, J. Eschner, M. Hennrich, G. Morigi, *Phys. Rev. A* **78**, 013808 (2008).
28. Q. M. Quan, I. Bulu, M. Loncar, *Phys. Rev. A* **80,** 011810 (2009).
29. G. –Y. Chen, N. Lambert, C-H. Chou, Y-N. Chen, F. Nori, *Phys. Rev. B* **84**, 045310 (2011).
30. J-J. Greffet, *Science* **308**, 1561 (2005).

**Supplementary Online Material**

**Sample preparation:** A $10^{-6}$ M solution of DBATT in tetradecane was prepared. 0.1 µL of this solution was applied onto a substrate containing patterned electrodes. The solid immersion lens was pressed on top and the whole sample was shock-frozen when inserted into the cryostat.

**Spectral instabilities:** In order to compensate for occasional small spectral fluctuations that may detune the source and target molecules during the measurement, we used an automated measurement protocol in which the frequencies of the two molecules were periodically matched. As a first step, fluorescence excitation spectra were recorded for both molecules in order to determine the Stark voltage $V_0$ that was required to bring them into resonance with each other. In the second step, the source molecule was pumped and the count rates on APD 4 were monitored as we offset the voltage by $\Delta V$. In practice, we scanned the Stark voltage in 32 steps of 50 ms residence times, so that a full scan lasted 1.6 s. We performed 80 scans after which we again determined $V_0$ and started a new series of scans. The data in Fig. 3 of the main manuscript resulted from 36 such cycles.

Figure (a) below shows an example of the collection of fluorescence excitation spectra of the target molecule recorded at the beginning of each of the 36 measurement cycles. At this point, it is not clear whether the observed drift of the resonance frequency in Fig. (a) is due to a frequency drift of the dye laser or of the target molecule. In order to adjust the spectral position of the source molecule to that of the target molecule, we tuned the laser frequency to the resonance of the target molecule (Fig. (a)) and excited the source molecule with the same laser beam. We then scanned its resonance frequency via the Stark effect and recorded a fluorescence excitation spectrum of the source molecule. The results are shown in Fig. (b) below and provide the voltages $V_0$ that need to be applied in each cycle to bring the two molecules into resonance with each other. Following this procedure, we ensured that the source and target molecule frequencies matched without having to know how much the laser frequency or the source frequency drifted.

In Fig. (c), we plot the amplitudes of the fluorescence excitation spectra from Fig. (a). We find a gradual decrease of the peak intensity caused by a slow spatial drift of the target sample with respect to the laser focus within about two hours. If we assume that the extinction dip of the target molecule (Fig. 2a in the main manuscript) follows the same trend throughout the measurement process, we can use this plot to assess a correction for the extinction dip in Fig 3. This leads to a correction of 8%, which we have implemented in the Lorentzian spectrum (red curve) in Fig. 3 of the manuscript accordingly.

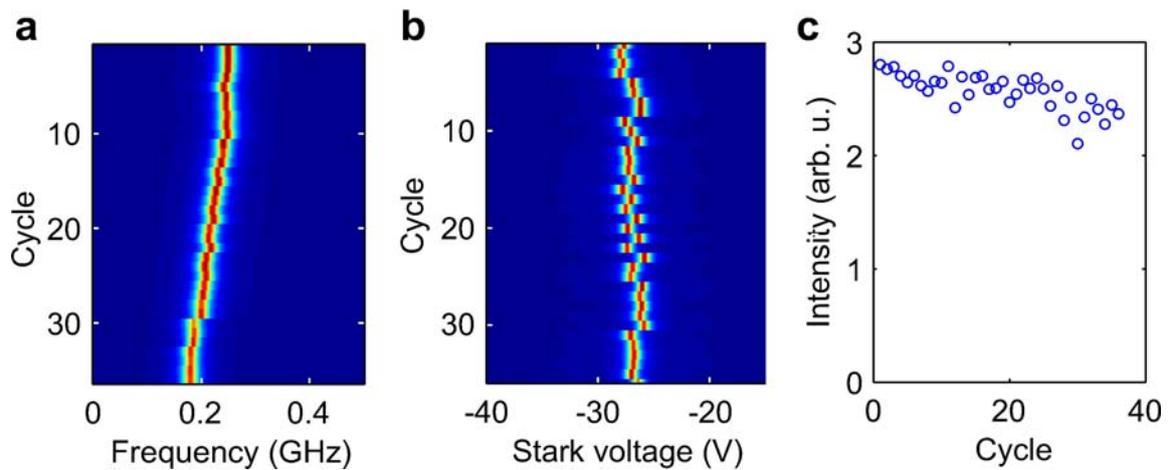

**Simulations:** The simulations of the photon wavepacket time evolution were based on a one-dimensional approach, where a Hamiltonian for a two-level system interacting with a photon was set up and the Schrödinger equation was solved within a single-photon basis set[1]. The simulations agreed with analytical calculations of the Wigner-Weisskopf type[2,3] within an accuracy limited solely by the finiteness of the basis set. We note that although the models in Refs. [1,2] are one dimensional, perfect reflection can also occur in three dimensions [4]. In Ref. [4] this is demonstrated for a directional dipolar wave that matches the radiation pattern of a dipolar emitter in free space. Another alternative is to use optical antenna such as discussed in Ref. [5] to adapt the radiation pattern of an emitter by engineering its electromagnetic environment.

**References:**

1 V. Buzek, G. Drobny, E. G. Kim, M. Havukainen, P. L. Knight, *Phys. Rev. A* **60**, 582 (1999).
2 U. Dorner, P. Zoller, *Phys. Rev. A* **66**, 023816 (2002).
3 S. Rist, J. Eschner, M. Hennrich, G. Morigi, *Phys. Rev. A* **78**, 013808 (2008).
4 G. Zumofen, N. M. Mojarad, V. Sandoghdar, M. Agio, *Phys. Rev. Lett.* **101**, 180404 (2008).
5 K. G. Lee, *et al.*, *Nature Photon.* **5**, 166 (2011).